# Co-Propagation of Quantum Time Synchronization and Optical Frequency Transfer over a 122 km Hollow-Core Fiber

Huibo Hong,[1,2,†] Xiao Xiang,[1,2,3,†] Runai Quan,[1,2,3,†] Rongduo Lu,[4] Qian Zhou,[1,2] Dawei Ge,[4] Liuyan Han,[4] Bo Liu,[1,2] Ru Yuan,[1,2] Dechao Zhang,[4] Yuting Liu,[1,2] Bingke Shi,[1,2] ZhiGuang Xia,[1,2] Xinghua Li,[1,2] Mingtao Cao,[1,2,3] Tao Liu,[1,2,3] Ruifang Dong,[1,2,3,*] and Shougang Zhang[1,2,3]

[1] *Key Laboratory of Time Reference and Applications, National Time Service Center, Chinese Academy of Sciences, Xi'an 710600, China*
[2] *School of Astronomy and Space Science, University of Chinese Academy of Sciences, Beijing, 100049, China*
[3] *Quantum Precision Measurement Department, Hefei National Laboratory, Hefei, 230026, China*
[4] *China Mobile Research Institute, Beijing 100053, China*
[†]*These authors contributed equally.*
* *dongruifang@ntsc.ac.cn*

**Abstract:** The co-propagation of quantum and classical signals through shared optical fibers is crucial for scalable quantum networks. However, this coexistence is fundamentally limited by spontaneous Raman scattering (SpRS) from the bright classical light, which generates overwhelming noise that disrupts the single-photon-level quantum signals. Here, we overcome this long-standing challenge by leveraging the inherently ultralow nonlinearity of hollow-core fiber (HCF) to suppress SpRS noise. By operating both the quantum time synchronization (QTS) and classical optical frequency transfer (OFT) signals within the telecom C-band, separated by only ~10 nm, we successfully demonstrate their simultaneous transmission over a 122-km HCF link. With a classical OFT power of 1 mW, the QTS performance shows negligible degradation, maintaining sub-picosecond time stability at 2000 s, while the OFT achieves a fractional frequency instability of $10^{-20}$. Near-sub-picosecond QTS stability is preserved even when the classical power is increased to 3 mW. Furthermore, simulations based on our experimental data indicate that with next-generation low-loss HCF, the platform can tolerate classical powers beyond 10 mW and extend the QTS range to over 500 km. By realizing a unified quantum–classical time–frequency distribution framework, this work establishes HCF as a highly capable and practical platform for future scalable quantum networks.

## 1. Introduction

The development of scalable quantum networks [1–3] hinges on the stable, long-distance co-propagation of quantum and classical signals within shared fiber infrastructure, which is critical from both a practical and economic standpoint. This integration enables core applications from quantum key distribution [4,5], distributed quantum sensing [6], to networked quantum computing [7,8] etc., while simultaneously boosting classical network security via quantum-assisted encryption [9,10]. The co-propagation, however, faces a fundamental bottleneck rooted in the extreme power disparity between bright classical signals and single-photon-level quantum signals. In standard silica fibers, this disparity triggers strong nonlinear crosstalk, notably via spontaneous Raman scattering (SpRS), that raises a spectrally broad noise floor capable of overwhelming the quantum signals [11,12].

Conventional mitigation of SpRS noise typically employs wide spectral separation between signals (e.g., one in the telecom C-band and the other in the O-band) [13–15], alongside wavelength-division multiplexing and narrowband filtering. While this strategy enables coexistence for specific quantum tasks [16,17], this approach introduces significant transmission loss to the signal operating outside the low-loss C-band window and remains fundamentally constrained by the intrinsic nonlinearity of silica fibers [18,19]. These limitations collectively result in either severely restricted co-propagation distances [20,21] or necessitate classical power to be reduced to the microwatt level [22]. Quantum entanglement paired with coincidence detection provides a signal-processing strategy. It isolates correlated photon pairs within a narrow temporal window to suppress accidental noise counts, thereby extending the feasible propagation distance [23]. Nevertheless, this strategy does not address the fundamental physical noise source. A more direct solution is to engineer the transmission medium. Guiding light in an air core, hollow-core fibers (HCF) inherently suppress the SpRS noise by orders of magnitude, tackling crosstalk at the physical level [24,25]. Beyond this core advantage, HCF also exhibits ultralow Rayleigh backscattering and thermal sensitivity [26,27], making it an ideal platform for stable and long-distance quantum-classical co-propagation.

Recent demonstrations have shown successful co-propagation of quantum signals with bright classical signals in HCF, as seen in Ref. [24,28,29]. Among the various services required for quantum networks, quantum time synchronization (QTS) and classical optical frequency transfer (OFT) represent two complementary paradigms. QTS harnesses quantum entanglement for secure and precise timing, whereas OFT exploits phase-coherent optical carriers for ultrastable frequency dissemination. However, integrating these two services within a single fiber presents a critical and still-unresolved challenge. To ensure co-propagation over long distances (>100 km), both signals must reside within the lowest-loss transmission window, which necessitates a moderate wavelength separation between them. Overcoming this challenge requires strictly balancing two competing demands: maintaining milliwatt-level classical power for OFT stability while suppressing the induced SpRS noise to preserve QTS performance. Successfully meeting this benchmark would not only validate the co-propagation paradigm but also shift it from passive coexistence to active mutual reinforcement. In such an integrated system, OFT could supply QTS with an ultrastable frequency reference[30–32], ensuring its long-term timing stability, while QTS could provide OFT a precise, quantum-secured timing reference[33–36] for advanced network applications such as distributed quantum sensing.

In this paper, we demonstrate the long-haul co-propagation of QTS and OFT over a 122-km HCF link, with both signals residing in the telecom C-band and separated by only 10 nm. With the OFT carrier operating at 1 mW, the QTS performance experiences negligible degradation, achieving a time deviation below 0.7 ps at 2000 s, while the OFT reaches a fractional frequency instability at the $10^{-20}$ level. The QTS performance maintains near-sub-picosecond even with 3 mW of co-propagating classical power. Beyond serving as a technical milestone for a unified quantum–classical time–frequency distribution system, this successful integration enables active mutual reinforcement between the two services. Furthermore, simulations based on our experimental data indicate a promising outlook for high-performance QTS, with a classical power tolerance exceeding 10 mW and a potential distance extension beyond 500 km using next-generation HCF. The demonstrated results and simulations provide us a precise, secure, and resilient metrological foundation for maintaining temporal coherence and enabling deterministic control across distributed quantum nodes, thereby supporting the scalable execution of advanced quantum communication, sensing, and computing protocols.

## 2. Experimental setup

The experimental setup for co-propagating QTS and OFT over HCF is shown in Fig. 1. The HCF, which interconnects the two separate sites (A and B), is a double-nested anti-resonant nanostructured fiber (DNANF) with a transmission loss of 0.17 dB/km. Since the QTS is implemented based on the quantum two-way time transfer (Q-TWTT) protocol [37], bidirectional transmission is employed. At each terminal, a custom-built Type-II spontaneous parametric down-conversion (SPDC) source [38] generates energy-time entangled photon pairs at 1561.79 nm with full width at half maximum bandwidth of 1 nm. For the source EPS-A

(EPS-B) at site A (B), the idler photons are kept locally, while the signal photons are multiplexed with the classical OFT carrier via a wavelength division multiplexer, WDM1 (WDM2), and routed through the HCF to the opposite site B (A). Both WDMs provide >120 dB isolation at 1550 nm to prevent classical light leakage into the quantum channel. Bidirectional routing is enabled by a couple of optical circulators (OC1 & OC2). The locally retained idler photons and the remotely received signal photons are individually detected by four single-photon detectors (denoted by SPD1-SPD4). In the experiment, free-running InGaAs SPDs (QCD600B, QuantumCTek Co., Ltd. ) are utilized, which features a quantum efficiency of 25%, a dark count rate of 5 kcps (kilo counts per second), and a timing jitter of about 200 ps. The arrival times of these detected photons are then stamped by the corresponding time taggers TTU-A and TTU-B (Time Tagger Ultra, Swabian Instruments) for coincidence measurement and time-difference extraction [39]. For experimental simplicity, both TTUs share a common Rb clock reference (Rock Electronic Ltd.). Note should be taken that, to cancel dispersive broadening of the photon pairs' temporal coincidence distribution width induced by the HCF, which features a group velocity dispersion of approximately 3.7 ps/nm/km, the nonlocal dispersion cancellation strategy [40] was employed by inserting a fiber Bragg grating-based dispersion compensation module (DCM, Proximion Inc.) in the local idler photon path (not shown in Fig. 1).

Parallel to the quantum channel, the OFT system utilized a narrow-linewidth continuous-wave (CW) laser (NKTE15, NKT Photonics) at 1550.07 nm. To suppress its sideband spectrum, the laser output was filtered by two cascaded C-band bandpass filters (BPF), providing >100 dB extinction at the operating band of QTS. The laser light is delivered from Site A to Site B via an OFT transmitter (OFT-TX) and received by an OFT receiver (OFT-RX), with both terminals incorporated an unbalanced interferometer. The beat note between the transmitted and received light extracts the fiber phase noise for active compensation, enabling ultrastable frequency transfer. A detailed OFT setup description can be found in Ref. [41].

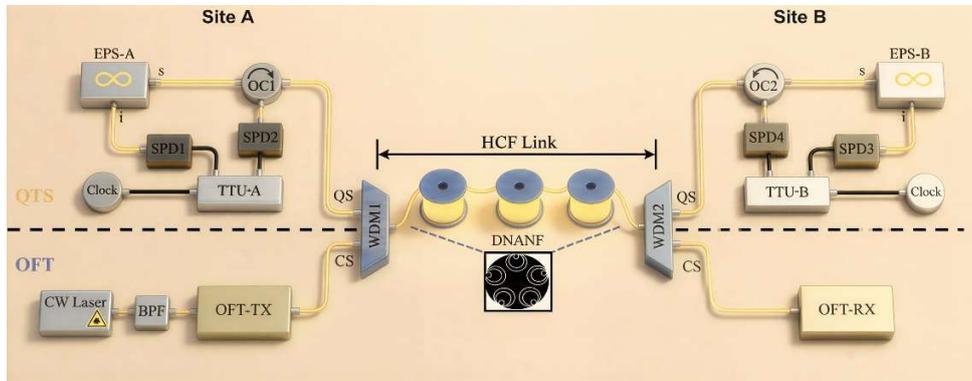

Fig. 1. Schematic diagram of the experimental setup for simultaneous QTS and OFT over HCF link. The system integrates a bidirectional QTS channel, based on energy-time entangled photon pairs (EPS-A & EPS-B), with an OFT channel. Both signals are combined and separated via high-isolation WDMs. EPS: Entangled Photon Source; OC: Optical Circulator; WDM: Wavelength Division Multiplexer; TTU: Time Tagger Ultra; BPF: Bandpass Filter; SPD: Single-Photon Detector.

## 3. Results and Analysis

### 3.1 Characterizing Quantum Signals under Co-Propagating Classical Light

To characterize the coexistence capability of QTS and OFT over HCF link, we removed the DCMs in the local idler paths and injected the CW laser directly into WDM1 after the cascaded BPFs. Subsequent power control was enabled by a variable optical attenuator. For power monitoring, a 90/10 fiber beam splitter was inserted, and 10% output was connected to an optical power meter, while the 90% output was combined with the quantum signal through WDM1. The coincidence-to-accidental ratio (CAR) was measured to quantify the signal-to-noise ratio of the quantum channel under co-propagating classical light. With both quantum and classical sources active, we characterized the CAR in the forward direction (A→B), accounting for the additional noise introduced by backward SpRS. By measuring the CAR as a function of classical carrier power (see Section S1, Supplemental Document for details), we extracted the HCF's intrinsic SpRS coefficient. The forward and backward SpRS coefficients of the HCF were measured to be 1.2 kcps/(mW·km) and 2.0 kcps/(mW·km) respectively, which were approximately two orders of magnitude lower than that of standard silica fibers [42], confirming the HCF's superior suitability for long-haul quantum-classical co-propagation.

To delineate the viable operating regime for simultaneous QTS and OFT in HCF, we modeled the system performance as a function of classical carrier power and HCF length (detailed derivation of the model is provided in Section S2, Supplemental Document). The resulting contour plot of CAR is presented in Fig. 2. In this plot, the color gradient corresponds to the CAR value, with warmer colors representing a higher CAR. This model was validated through two targeted experimental scans, whose results are overlaid as discrete data points in Fig. 2. Both scans show excellent agreement with the theoretical predictions. In the power-dependence scan (circular data points), conducted at a fixed length of 54 km (vertical dashed line), sweeping the classical power from 0.5 mW to 3.0 mW revealed the theoretically predicted monotonic decrease in CAR due to SpRS noise. Crucially, even at 3.0 mW, the measured CAR remained above 10 [20], affirming HCF's strong SpRS noise suppression. In the distance-dependence scan (square data points), performed at a fixed classical optical power of 1 mW (horizontal dashed line), increasing the HCF length up to 213 km yielded CAR values governed by signal attenuation and noise accumulation. The combined approach of direct noise measurement and theoretical modeling provides a clear guideline for a unified quantum-classical time-frequency network based on HCF.

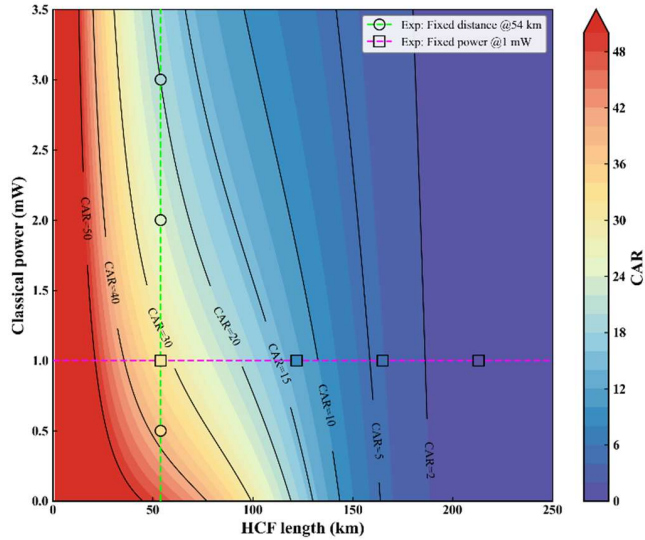

Fig. 2. Experimental and theoretical characterization of quantum signal quality under classical co-propagation. The contour plot shows the simulated CAR as a function of fiber length and classical carrier power based on utilized HCF parameters. Circles denote experimental data acquired at a fixed distance of 54 km while varying the classical power (green dashed line). Squares denote data acquired at a fixed classical power of 1 mW while varying the fiber length (magenta dashed line).

*3.2 QTS Performance with Co-Propagating OFT*

The time deviation (TDEV) performance of QTS under co-propagation with a 1 mW OFT signal is plotted in Fig. 3 (green squares). For comparison, stability curves for representative commercial frequency standards are also shown: a Rubidium (Rb) atomic clock (magenta triangles) and a high-performance Cesium (Cs) beam atomic clock [43] (orange pentagons). The TDEV of QTS exhibits a characteristic white phase noise behavior and surpasses the stability of both the Rb and Cs clocks across the entire measurement duration from 20 s to 2000 s. At an averaging time of 2000 s, the TDEV reaches 0.68 ps, maintaining sub-picosecond-level performance consistent with the baseline obtained without OFT co-propagation (0 mW, denoted by blue circles in Fig. 3). This convergence confirms that the residual noise induced by the 1 mW classical laser is negligible, demonstrating the superior SpRS noise suppression capability of the HCF platform.

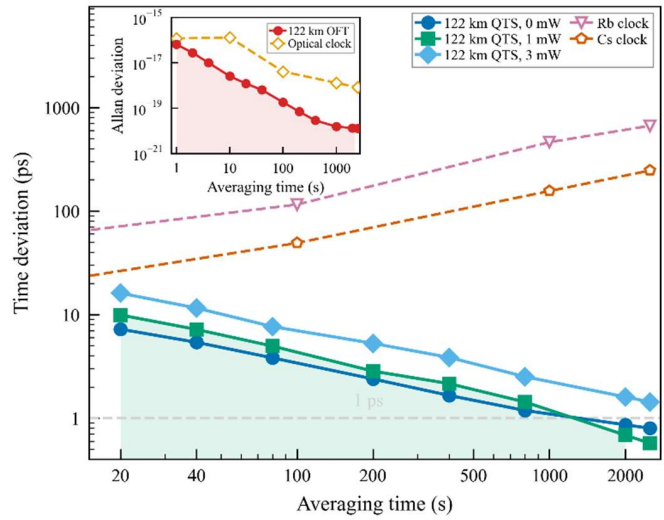

Fig. 3. Stability of co-propagating QTS and OFT over 122 km HCF link. The TDEV of QTS is shown under 0 mW (blue circles), 1 mW (green squares), and 3 mW (cyan diamonds) OFT carrier powers. Dashed lines indicate the stability of commercial Rubidium (Rb, magenta triangles) and Cesium (Cs, orange pentagons) frequency standards for reference. The inset displays the Allan deviation of the OFT (red circles) compared with the stability of an optical clock (orange diamonds).

When the OFT carrier power is increased to 3 mW, the corresponding QTS stability is plotted as cyan diamonds in Fig. 3. An upward shift in the stability curve is discerned while the system maintains near-sub-picosecond stability at an averaging time of 2000 s. We attribute this degradation mainly to saturation of the SPDs caused by the increased flux of SpRS noise photons (see Supplemental Document, Section S2.3). The increased noise photon flux causes more genuine signal photons to fall within the SPD's dead time (~1 μs), reducing the effective coincidence detection rate. This effect directly translates into the degraded timing stability and could be mitigated by using SPDs with a high saturation count rate [44]. Despite this effect, the stability curve maintains its $\tau^{-1/2}$ scaling, confirming the inherent robustness of the system and its tolerance to higher-power classical co-propagation.

The fractional frequency instability of the OFT system over the 122-km HCF link, evaluated in terms of Allan deviation, is shown in the inset of Fig. 3. The system achieves an instability of $3\times10^{-17}$ at 1 s, which scales down to $1.3\times10^{-20}$ at 2000 s. This performance is better than that of state-of-the-art optical clocks [45] (orange diamonds in the inset) and confirms that the OFT system, operating at the 1 mW power level used for quantum co-propagation, meets the stringent stability required for optical clock comparisons. Taken together, these results demonstrate that the HCF-based platform can deliver both ultrastable frequency transfer and high-precision quantum time synchronization within a single fiber, thereby establishing a hybrid quantum-classical time–frequency distribution foundation for future advanced quantum networks.

## 4. Discussion

The presented results demonstrate that 122 km hollow-core fiber platform enables high-fidelity co-propagation of quantum time synchronization and optical frequency transfer, with both subsystems meeting the stringent requirements of future hybrid quantum-classical time-frequency networks. The key performance metrics of quantum-classical co-propagation in fibers are summarized in Table 1, focusing on polarization [17,21,28] and energy-time [20,22] as the two entanglement degrees of freedom (DOF). By leveraging the inherent advantages of HCF, this work extends the coexistence distance of entanglement and classical light beyond the previous distance record of 100 km in standard SMF [17], while supporting an improved classical optical power more than 1000 times (30 dB). A primary motivation for implementing such high-power classical co-propagation is that, unlike conventional data transmission which operates at a relatively low optical power, the classical signal here serves as the optical carrier itself for coherent transmission. Correspondingly, this elevated power also enables higher data transmission rates. Furthermore, by operating both classical and quantum signals within the C-band with wavelength separation of 11.7 nm, our platform ensures simultaneous low-loss transmission for both services.

**Table 1 Performance metrics for quantum and classical signals co-propagating in standard and hollow-core optical fibers**

| Ref. | Fiber type & length (km) | Optical power (dBm) | Wavelength separation (nm) | Quantum/Classical wavelength allocation | Entanglement DOF |
|---|---|---|---|---|---|
| *This work* | HCF, 122 | +4.7 | 11.7 | C-band/C-band | Energy-time |
| [28] | HCF, 11.5 | -3 | 13.7 | C-band/C-band | Polarization |
| [17] | SMF, 100 | -26[a] | 267.4 | C-band/O-band | Polarization |
| [21] | SMF, 60 | +4 | 236.6 | O-band/C-band | Polarization |
| [20] | SMF, 40 | -16[a] | 16.2 | C-band/L-band | Energy-time |
| [22] | SMF, 40 | -13[a] | 7.1 | C-band/C-band | Energy-time |

[a]These values are derived based on the received optical power and corresponding link loss.

To further explore the potential of quantum-classical co-propagation in HCF, we simulated the achievable CAR as a function of transmission distance using the validated framework in Supplemental Document S2. As presented in Fig. 4, the analysis models the state-of-the-art scenario based on the next-generation HCF with a transmission attenuation of 0.08 dB/km [46,47] and advanced superconducting nanowire single-photon detectors (SNSPD, 90% detection efficiency and a dark count rate of 50 cps) [48,49]. This optimized configuration enables a flexible trade-off between transmission distance and classical signal power. With a co-propagating classical signal power of 1 mW, the system can be extended to a distance of 563 km (orange dashed line), which is sufficient for metropolitan-area quantum networks without the need for quantum repeater [50] or cascading strategies [51]. At the other end, even at a high classical power of 10 mW, the system remains operational over 138 km (purple solid line). Notably, this distance is comparable to that achieved in our current experiment, yet it supports an order-of-magnitude higher classical power. This capability significantly broadens the range of compatible classical services and multi-node application scenarios.

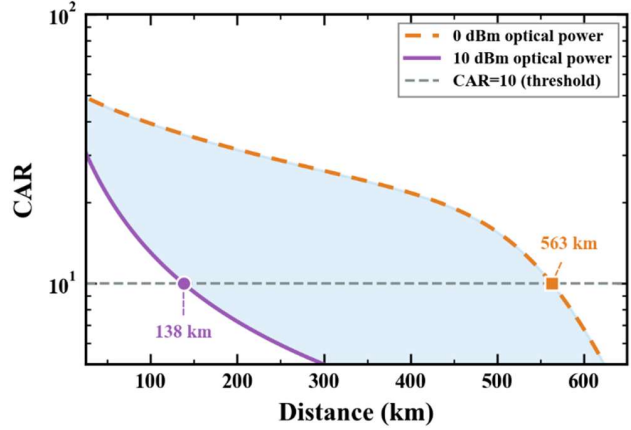

Fig. 4. Simulated quantum-classical coexistence performance for the state-of-the-art scenario integrating next-generation hollow-core fiber (0.08 dB/km attenuation) and advanced superconducting nanowire single-photon detectors (90% efficiency, 50 cps dark count rate). The orange dashed and purple solid lines represent the CAR with co-propagating optical power of 0 dBm and 10 dBm, respectively.

## 5. Conclusion

Fundamentally addressing the challenge of SpRS noise, this work demonstrates HCF as a practical platform for hybrid quantum–classical networks. The ultralow nonlinearity of HCF suppresses the SpRS noise that typically disrupts quantum signals, transforming the classical channel from a source of interference into a compatible partner for high-fidelity co-propagation.

We have established and experimentally validated a unified quantum-classical time-frequency distribution platform over a 122-km HCF link. Operating both quantum and classical signals within a single telecom band (C-band) with only 10-nm spectral spacing, this platform simultaneously delivers time synchronization with sub-picosecond stability and optical frequency transfer with $10^{-20}$ level instability. The system maintains this performance even at a classical power of 3 mW, demonstrating exceptional resilience and practical utility. These results confirm that HCF facilitates not just passive coexistence but active, synergistic operation between quantum and classical channels. By providing a noise-resilient, high-capacity foundation for unified time-frequency distribution, our work lays the groundwork for scalable and functionally integrated quantum systems. It thereby directly advances secure communication, distributed quantum sensing, and large-scale quantum computing, marking a decisive step toward a coherent hybrid quantum–classical infrastructure.

Building on the experimental results, we further establish a scalable performance roadmap for this platform. By integrating measured parameters with realistic projections for next-generation low-loss HCF, our analysis shows that the system can tolerate classical powers exceeding 10 mW over a fiber length of 138 km while capable of extending the direct quantum transmission distance beyond 500 km. This scalability not only validates the technical viability for metropolitan and regional quantum networks but also highlights the transformative role of HCF in building practical, large-scale quantum infrastructures.

**Funding.** National Natural Science Foundation of China (Grant Nos. 12033007, 12103058, 12203058, 12074309), Youth Innovation Promotion Association of the Chinese Academy of Sciences (2021408, 2022413, 2023425), and the Innovation Program for Quantum Science and Technology (2021ZD0300900).

**Author Contributions.** H.H., X.X., and R.Q. contributed equally to this work. H.H. performed the experiments, analyzed the data, and wrote the initial draft. X.X. and R.Q. contributed to experimental design, conceptualization, and manuscript optimization. Q.Z., B.L., and R.Y. constructed the OFT system. Y.L., B.S., and Z.X. assisted in QTS experiments. R.L., D.G., L.H., and D.Z. provided the HCF infrastructure and site support. X.L. and M.C. provided conceptual guidance. T.L. supervised the OFT work. R.D. supervised the project as principal investigator, provided conceptual guidance, and revised the manuscript. S.Z. provided overall supervision and institutional support. All authors discussed the results and approved the final manuscript.

**Disclosures.** The authors declare no competing interests.

**Data availability.** All of the data that support the findings of this study are reported in the main text and Supplemental Document. Source data are available from the corresponding authors on reasonable request.

**Supplemental document.** See Supplemental Document for supporting content.